\documentclass{llncs}
\usepackage[latin1]{inputenc}
\usepackage[english]{babel}
\usepackage{amsmath}
\usepackage{amsfonts}
\usepackage{amssymb}
\usepackage{graphicx}
\usepackage{subfigure}
\usepackage{tikz}

\usetikzlibrary{matrix}

\author{Alexander L\"uck\inst{1} \and Verena Wolf\inst{1}}
\institute{Department of Computer Science, Saarland University, Saarbr\"ucken, Germany
}

\title{Generalized Method of Moments Estimation for Stochastic Models of DNA Methylation Patterns}
\begin{document}
\maketitle

\begin{abstract}
With   recent advances in  sequencing technologies, large amounts of epigenomic data have become available and computational methods are  contributing significantly to the progress of epigenetic research. As an orthogonal approach to methods based on machine learning, mechanistic modeling aims at a description of the mechanisms underlying epigenetic changes.
Here, we propose an efficient method for parameter estimation for stochastic models that describe the dynamics of DNA methylation patterns over time.
Our method is based on the Generalized Method of Moments (GMM) and 
gives results with an accuracy similar to that of  maximum likelihood-based estimation approaches. 
However, in contrast to the latter,  the GMM still allows an efficient and accurate calibration of parameters even if
the complexity of the model is increased by considering longer methylation patterns. 
 We show the usefulness of our method by applying it to  hairpin bisulfite sequencing data from mouse ESCs for varying pattern lengths. 
\end{abstract}

\section{Introduction}
Epigenetics is an emerging field that is concerned with the study of heritable changes in the regulation of gene expression that are not a result of changes in the DNA sequence.
Epigenetic mechanisms, such as DNA methylation and histone modifications, can change the chromatin structure, regulate gene expression, and control cellular development and differentiation in higher organisms.
Epigenetic marks are also increasingly being recognized as important elements underlying  diseases such as cancer and certain autoimmune, neurodegenerative, as well as psychological disorders.

With the rapid evolution of high-throughput technologies
 for epigenetic analysis,  data on a genome-wide scale is available \cite{booth2012quantitative,booth2013oxidative,giehr2018two,laird2004hairpin,lutsik2011biq} and computational methods are  contributing significantly to the progress of epigenetic research. 
For instance, deep learning can be used to impute the methylation state at individual DNA positions if information about the state of neighboring positions is available 
\cite{angermueller2017deepcpg}.
As an orthogonal approach to   learning-based methods, which focus on accurate predictions, mechanistic models have been developed to describe the mechanisms underlying epigenetic changes and test different hypotheses \cite{Finch1897,genereux2005population,otto1990dna,sontag2006dynamics}.

Arand et al. proposed a Hidden Markov model (HMM) for the evolution of DNA methylation patterns during early development and applies it to hairpin bisulfite sequencing data from mouse
 embryonic stem cells \cite{arand2012vivo}. 
It gives a mechanistic description of the activity of the DNA methyltransferases Dnmt1, Dnmt3a, and Dnmt3b over time, as well as the loss of methyl groups through cell division.
Since in mammals, DNA methylation primary occurs   on the cytosine nucleotide of a CpG site, the model considers the methylation state of individual CpGs over time.  
Trained on KO data, the model is able to predict unseen methylation patterns in wild-type.
A similar model has been used to gain insights into the detailed molecular mechanisms underlying passive and active demethylatyion \cite{kyriakopoulos2019hybrid}.
Moreover, for genome-wide data, parameter values that describe the efficiency of epigenetic modifications 
in such models can be clustered  and correlated with data from enrichment analysis  \cite{kyriakopoulos2019stochastic,luck2019hidden}.

Several mechanistic models have been proposed that
consider methylation patterns of a number of successive 
CpGs and their spatial relationships \cite{bonello2013bayesian,lacey2009modeling,meyer2017modeling}.
Here, we consider a spatial extension of the models considered in \cite{arand2012vivo,kyriakopoulos2019hybrid},
which has been proposed recently \cite{luck2019hidden,luck2017stochastic}.
Its main strength compared to other models is that for each locus, it 
considers methylation efficiencies and dependency parameters. 
Moreover, it describes the methylation state of both DNA strands and 
is thus appropriate for data from hairpin bisulfite sequencing \cite{giehr2018hairpin}.

A major challenge   is that the complexity of models considering methylation patterns of several CpGs is much higher than the complexity of models that consider a 
 CpG in isolation. In the former case, all possible combinations
 of states of the individual CpGs have to be considered during the analysis.
Standard numerical approaches for parameter estimation based on maximizing the  likelihood of the data \cite{arand2012vivo} fail for such models, since the number of possible states 
is too large. 
Likelihood-free approaches based on stochastic sampling, such as   Approximate Bayesian Computation have been applied in this context \cite{bonello2013bayesian}.
They allow to  estimate the posterior distribution based
on a comparison of  measured and  simulated data sets
  but often suffer from slow convergence to the true
posterior distribution. 

Here, we propose an approach that is not based on sampling but exploits the regular structure of the underlying Markov model. 
We suggest a number of statistical moments of the model that are most informative for calibration. Then we
 use a Generalized Method of Moments (GMM) estimator that considers weighted differences between the   moments estimated from the data and 
 those of the model.  
The GMM approach is a very popular likelihood-free technique  that was originally developed in econometrics \cite{hall2005generalized,wooldridge2001applications}. 
Its main advantage is that only the statistical moments of the model have to be computed but not the full underlying 
probability distribution. 
In the case that equations for the evolution of the statistical moments are available, a very fast estimation is possible,
which becomes more accurate when the order  of the considered moments is increased  \cite{backenkohler2016generalized,luck2016generalized}.
If the models'  moment equations are not available, the moment values can still be efficiently estimated through stochastic sampling.
We use the GMM for estimating methylation efficiencies and parameters that describe the dependence between neighboring CpGs.  
We determine a number of statistical moments that are most informative for 
identifying these parameters and compare our results to that of maximum likelihood estimation for short patterns, where a full numerical solution is possible.
After evaluating the accuracy on artificial data, we apply our approach to data from   hairpin bisulfite sequencing of mouse ESCs, where we used the same data as in \cite{arand2012vivo}.

This paper is organized as follows:
In Section~2 we briefly describe the methylation model and introduce the GMM framework. The results are presented in Section~3 and we conclude our findings in Section~4.
\section{Methods}
\subsection{Model}
We consider  a spatial stochastic model that describes the evolution of methylation patterns on double stranded DNA methylation data over time \cite{luck2017stochastic}.
As each CpG contains two Cs (one on each strand), each of which can either be methylated or unmethylated, there are
  four possible states for a CpG: both Cs unmethylated (state 0), only the C on the upper strand is methylated (state 1), only the C on the lower strand is methylated (state 2) or both Cs methylated (state 3).
A methylation pattern can be considered as concatenation of these methylation states.
Fig.~\ref{Fig:states} shows an example of pattern 0123.    
State transitions may occur due to cell division, maintenance and \emph{de novo} methylation.
During cell division one strand and its methylation state is kept as it is (parental strand) and the other strand is newly synthesized (daughter strand) initially containing  only unmethylated Cs. 
Then, with probability $\mu$,  an unmethylated C on the daughter strand is methylated through maintenance methylation, if the C on the parental strand is already methylated.
Moreover, with probability $\tau$  \emph{de novo} methylation may happen on unmethylated Cs on both strands, independent of the methylation state of the other strand.
This simple model for a single CpG   defines   a Discrete-Time Markov Chain \cite{arand2012vivo}.
To describe methylation patterns, i.e.,   sequences of $L$ CpGs with  $4^L$ possible states, we consider an extended model for multiple CpGs, which 
incorporates    additional parameters $\psi_L$ and $\psi_R$ for the dependency to the left and to the right \cite{luck2017stochastic}.
Intuitively, the higher $\psi_L$ ($\psi_R$) the more independent is 
the probability of being methylated 
of the methylation state of the left (right) neighbor, respectively.
Data from ESCs with
Dnmt3a/b DKO, for instance, gives
$\psi_L\approx \psi_R\approx 1$
when fitted to this model because 
maintenance through Dnmt1 occurs independent of the state of neighboring
CpGs. In contrast, calibration of the model to Dnmt1 KO data shows a clear dependence of the methylation activity of Dnmt3a/b to the left \cite{luck2019hidden}. 

The transition probability matrices
of the  model for $L$ CpGs can be generated based on the matrices of a single CpG using a Stochastic Automata Network approach with functional transitions \cite{alex2019stochastic}. 
These functional transitions take into account the state of neighboring CpGs through the dependency parameters $\psi_L$ and $\psi_R$.
For more details about the model we refer to \cite{luck2017stochastic,alex2019stochastic}.

\subsection{Generalized Method of Moments}
The main idea of moment-based parameter estimation methods is to directly compare certain theoretical moments of the model and the corresponding sample moments of the data (method of moments) or to minimize a score function based on theoretical and sample moments (generalized method of moments; GMM).
To ensure identification of the parameters, we use the following quantities, which are   based on the methylation state and independent of the labeling of these states.

\begin{figure}[tb]
\begin{center}
\includegraphics[width=0.95\textwidth]{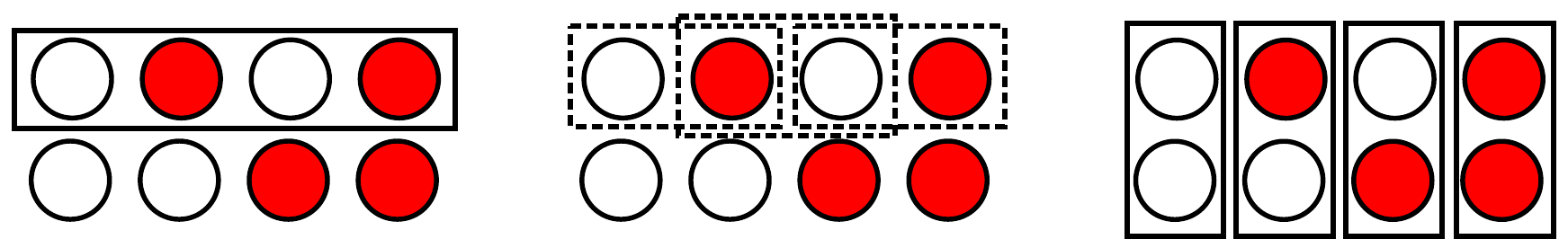}
\caption{Visual representation of the random variables \eqref{eq:methlevel} (left), \eqref{eq:proc1}, \eqref{eq:proc0} (middle), and \eqref{eq:methnumber} (right) for 4 CpGs and example pattern 0123. When only considering the upper strand, this pattern is converted to 0101. \eqref{eq:varlevel} and \eqref{eq:varnumber} correspond to the variances of \eqref{eq:methlevel} and \eqref{eq:methnumber}.  \label{Fig:states}}
\end{center}
\end{figure}

We consider a pattern of $L$ CpGs in the $k$-th measured cell, $k \in \{1,\ldots, N\}$. Let $M_i^{(k)}\in\{0,1\}$ be the methylation state of the upper C in CpG $i$, where 1 represents a methylated and 0 an unmethylated C.
Let $S_i^{(k)}\in \{0,1,2\}$ be the number of methylated Cs of CpG $i \in \{1,\ldots, L\}$.
We consider moments of the following random variables:
\begin{itemize}
\item (the horizontal average of) the  methylation level on the upper strand  
\begin{equation}
X_k=\frac{1}{L}\sum_{i=1}^L M_i^{(k)},
\label{eq:methlevel}
\end{equation}
\item the squared difference of the methylation level and the (cell population) average of the  methylation level  
\begin{equation}
(X_k-\bar{X})^2,
\label{eq:varlevel}
\end{equation}
\item a quantity to measure the fraction of consecutive methylated Cs on the upper strand
\begin{equation}
\frac{1}{L-1}\sum_{i=1}^{L-1} \left(M_i^{(k)}\cdot M_{i+1}^{(k)}\right),
\label{eq:proc1}
\end{equation}
\item a quantity to measure the fraction of consecutive unmethylated Cs on the upper strand
\begin{equation}
\frac{1}{L-1}\sum_{i=1}^{L-1} \left((1-M_i^{(k)})\cdot (1-M_{i+1}^{(k)})\right),
\label{eq:proc0}
\end{equation}
\item the   number of methylated Cs for each CpG
\begin{equation}
S_i^{(k)},
\label{eq:methnumber}
\end{equation}
\item the squared difference of the number of methylated Cs and the cell population average   of methylated Cs in each CpG
\begin{equation}
(S_i^{(k)}-\bar{S_i})^2
\label{eq:varnumber}
\end{equation}
\end{itemize}
Note that since our model is strand symmetric, the upper and lower strand behave equivalently and the moments based on Eqs.~\eqref{eq:methlevel}-\eqref{eq:proc0} are identical for both strands.
Therefore,  w.l.o.g. we consider only the quantities for the upper strand.
A visual representation of the quantities can be found in Fig.~\ref{Fig:states}.

We selected the above quantities  based on some considerations: 
The methylation level \eqref{eq:methlevel} and number of methylated Cs for each CpG \eqref{eq:methnumber} are obvious choices.
The squared differences to their average \eqref{eq:varlevel} and \eqref{eq:varnumber} are later needed to obtain variances.
Since the model contains neighborhood dependencies, i.e., the state of one CpG may influence (or even determine) the states of its neighbors, the number of consecutive (un)methylated Cs \eqref{eq:proc1} and \eqref{eq:proc0} contain valuable information.
Note that with only one of these quantities, it is not possible to distingush between alternating states and consecutive opposite states, e.g. with  Eq.~\eqref{eq:proc1} only, it is impossible to distinguish the patterns $00000$ and $10101$ ($L=5$).
The combination of \eqref{eq:proc1} and \eqref{eq:proc0} contains this information.
We investigate, which of the defined quantities \eqref{eq:methlevel}-\eqref{eq:varnumber} are mandatory for the successful parameter identification and estimation.

For each measured cell $k$, we collect the quantities \eqref{eq:methlevel}-\eqref{eq:varnumber} (or a subset thereof) in a random vector~${\bf{Y}}_k$.
For $L$ CpGs, each ${\bf{Y}}_k$ has (depending on how many moments are used, up to) $m=4+2L$ entries.
The corresponding sample moments are denoted by
\begin{equation}
\overline{\bf{Y}}=\frac{1}{N}\sum_{k=1}^N \vec{Y}_k
\end{equation}
and the theoretical moments, which can be obtained from the numerical solution of the model, are denoted by $\vec{m}(\theta)$, where $\theta$ is the vector of model parameters. 
We define the cost function  
\begin{equation}
{\bf g}(\theta):=\overline{\bf{Y}}-\vec{m}(\theta).
\label{eq:cost}
\end{equation}  
Since the entries of ${\bf g}(\theta)$ may in general be correlated, we define a class of estimators that also take mixed terms between the entries into account. 
Let $W$ be a positive semi-definite $m\times m$ matrix.
The GMM estimator 
\begin{eqnarray}\label{eq:gmm}
\hat{\theta} &=&\arg\min_\theta  {\bf g}(\theta)' W {\bf g}(\theta),
\end{eqnarray}
 was originally introduced by Hansen \cite{hansen1982large}.
Note that for $W=I$ (identity matrix) Eq.~\eqref{eq:gmm} corresponds to the least-squares estimator with $m$ terms.
For a general $W$ the number of terms in the cost function increases to $\tilde{m}=\frac{m(m+1)}{2}$.
Note that in order to identify the parameters, we need at least as many constraints in the cost functions as there are unknown parameters.
Including the mixed terms by choosing 
non-zero values for the off-diagonal entries in $W$  increases the number of constraints, which is often beneficial.
We assume consistency, i.e.,
$$\text{E}[\vec{Y}]=\vec{m}(\theta)~\text{if and only if } \theta=\theta_0,$$
where $\theta_0$ is the true parameter set. Then, it can be shown that choosing $W=F^{-1}$, with
\begin{equation}
F=cov[\vec{Y},\vec{Y}]
\label{eq:cov}
\end{equation}
yields an estimator with   smallest variance \cite{hall2005generalized,hansen1982large}.
Intuitively, whenever a sample moment
has high variance, its weight is decreased compared to sample moments with lower variance.
Since the covariance depends on the (unknown) real parameters $\theta_0$, one can use a multistep approach, starting with $W=I$ and iteratively reestimate a value $\tilde{\theta}$ in order to improve the estimation of $\theta$.
However, using the estimated $\tilde{\theta}$ may lead to misspecification
$$\text{E}[\vec{Y}]\neq\vec{m}(\tilde{\theta}),$$
such that the weight matrix may not be ideal.
Another approach is the so-called \emph{demean estimator}, where the sample counterpart of Eq.~\eqref{eq:cov}, i.e., 
\begin{equation}\label{eq:hatF}
 \hat F=\frac{1}{N}\sum^N_{k=1} ({\bf Y}_k - \overline{\bf Y})({\bf Y}_k - \overline{\bf Y})^T
 \end{equation} 
is used to resolve this inconsistency \cite{hall2005generalized}.
For the remainder of this paper, we will focus on results from the demean estimator and denote the corresponding estimator by $\hat\theta_{\mbox{\tiny GMM}}$.

\section{Results}
In order to determine the accuracy of the GMM approach applied to parameters of spatial methylation models, we initially use artificial data (with known parameters) generated from   Monte-Carlo (MC) simulations.
Additionally, we compare the GMM estimations to results from a Maximum Likelihood Estimation (MLE)
\begin{equation}
\hat{\theta}_{\mbox{\tiny MLE}}= \arg \max_{\theta} \ell (\theta), \quad \ell(\theta)=\sum_{j=1}^{4^L} \log(\hat{\pi}_j(\theta))\cdot N_j,
\label{eq:MLE}
\end{equation}
where $\hat{\pi}_j(\theta)$ is the probability of pattern with index $j$ obtained from the numerical  solution of the model for parameters $\theta$ and $N_j$ the observed count of the $j$-th pattern from the MC simulations.
To enumarate the patterns, each pattern can be considered as a number in the tetral system, which can be converted to the decimal system in order to obtain the unique index $j$.
For each parameter set and sample size we generate $25$ data sets from MC simulations and use them to obtain the mean and standard deviations for the estimates.

\begin{figure}[tb]
\begin{subfigure}[True parameters: $\mu=0.8$, $\psi_L=0.4$]{\includegraphics[width=0.49\textwidth]{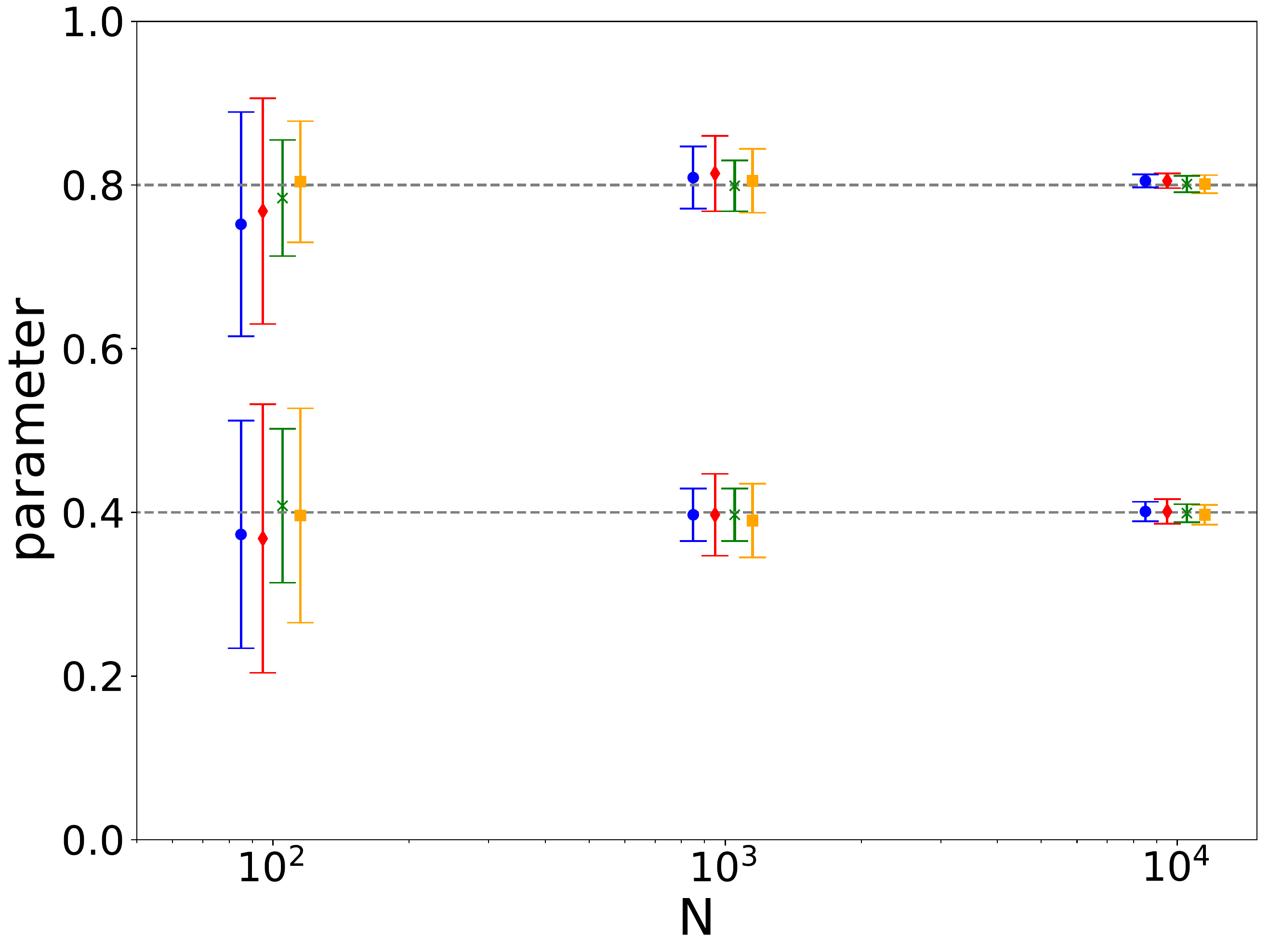}}
\end{subfigure}\hfill
\begin{subfigure}[True parameters: $\tau=0.1$, $\psi_R=0.6$]{\includegraphics[width=0.49\textwidth]{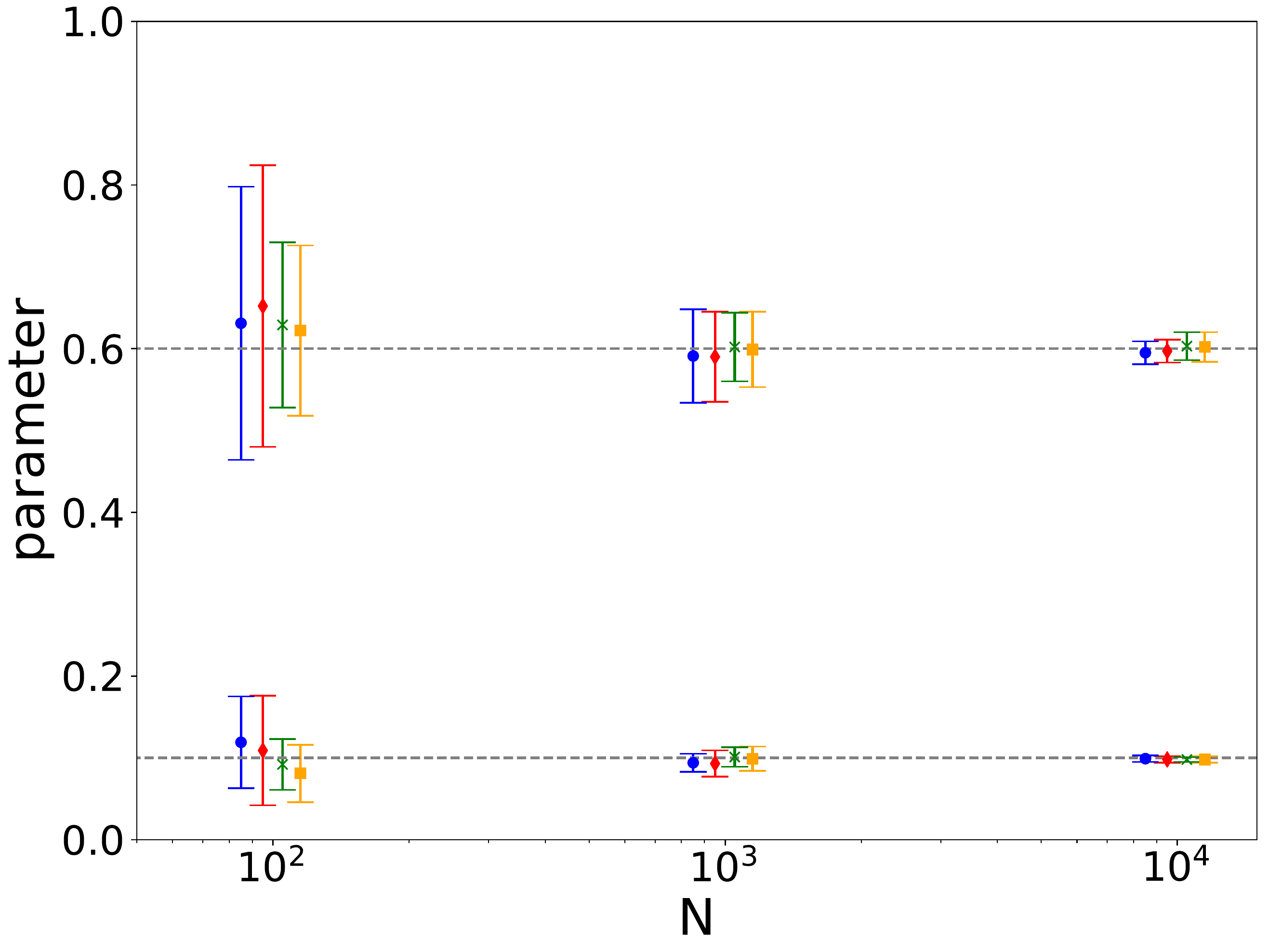}}
\end{subfigure}\\[-3ex]
\caption{Mean and standard deviation of the estimated parameters $\hat{\theta}_{\mbox{\tiny GMM}}$ and $\hat{\theta}_{\mbox{\tiny MLE}}$ from 25 estimations for MC simulation data with a sample size of $N$. The red (orange) bars show the GMM estimations for 3 (4) CpGs and the blue (green) bars the MLE estimations for 3 (4) CpGs.\label{Fig:compGMMMLE}}
\end{figure}

In Fig.~\ref{Fig:compGMMMLE} we plot the
  results for parameters $\theta=(\mu,\psi_L,\psi_R,\tau)=(0.8,0.4,0.6,$ $0.1)$, where the red (orange) bars show the GMM estimations for $L=3$ ($L=4$) CpGs and the blue (green) bars the MLE estimations for $L=3$ ($L=4$) CpGs for different sample sizes $N$, respectively.
  Note that we assume identical parameters for all CpGs of the pattern and $N$ is the number of single-cell pattern samples at the selected position.
Also note that the bars have a little offset to the left/right of the actual sample size in order to increase the clarity of the presentation.
We observe that both GMM and MLE show a very similar performance in terms of accuracy for all four parameters.
Furthermore, a relatively modest sample size of $100-1000$  is already enough to obtain reliable estimates.

\begin{figure}[tb]
\begin{center}
\includegraphics[width=0.7\textwidth]{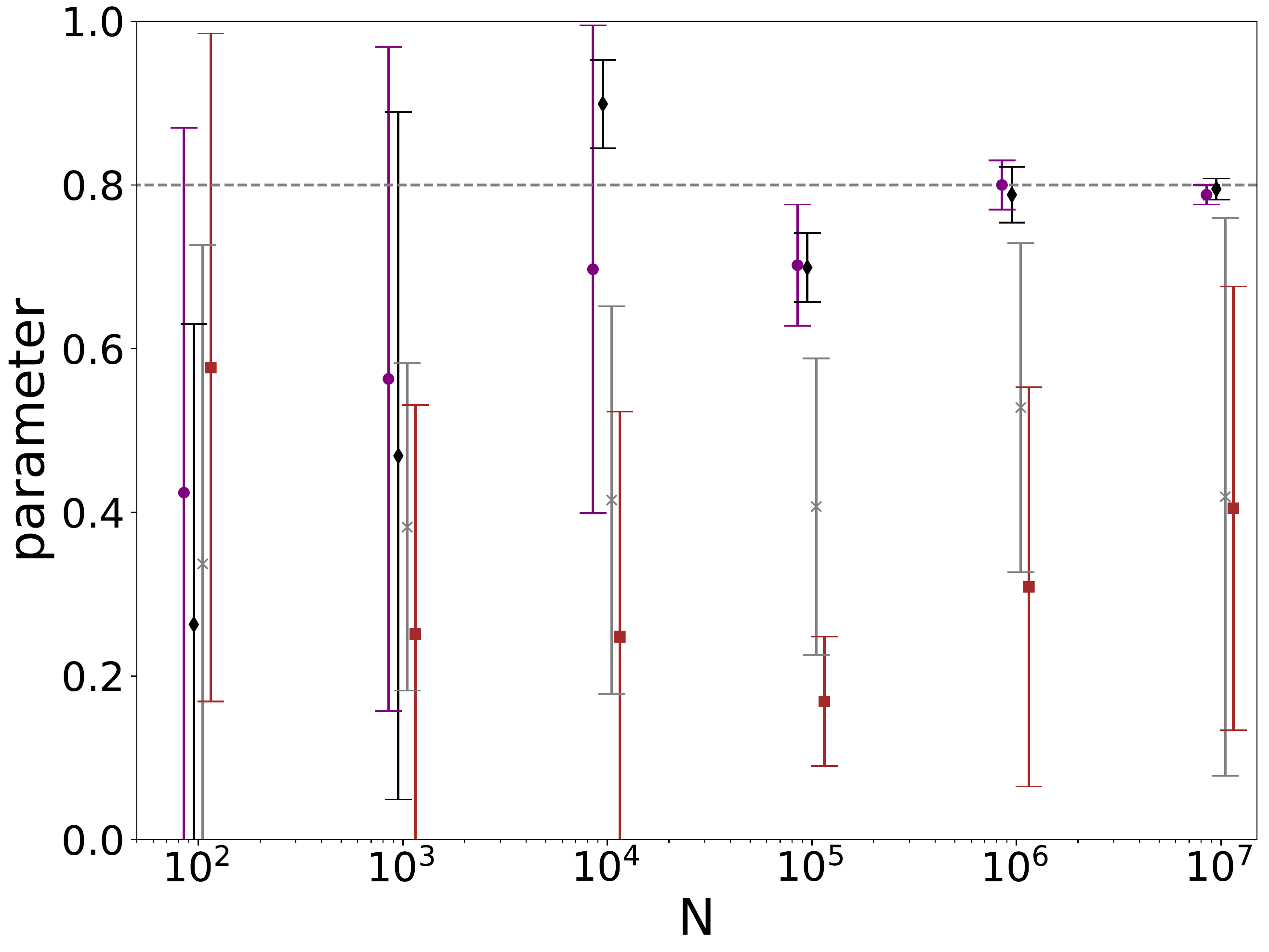}
\caption{Estimations for $\mu$ for different subsets of moments. Purple: \eqref{eq:proc1}, \eqref{eq:methnumber}; black: \eqref{eq:methlevel}, \eqref{eq:proc1}, \eqref{eq:proc0}, \eqref{eq:methnumber}; gray: \eqref{eq:methlevel}, \eqref{eq:proc1}, \eqref{eq:proc0}; brown: \eqref{eq:methlevel}, \eqref{eq:methnumber} \label{Fig:lessmoms}}
\end{center}
\end{figure}
Note that not all moments derived from Eqs. \eqref{eq:methlevel}-\eqref{eq:varnumber} are needed to ensure identifiability of the parameters.
In Fig. \ref{Fig:lessmoms}, we plot 
results for different subsets of moments.
Without the variances (Fig. \ref{Fig:lessmoms}, black bars, moments of Eqs. \eqref{eq:methlevel}, \eqref{eq:proc1}, \eqref{eq:proc0}, \eqref{eq:methnumber}) and additionally even without the methylation level and the successive unmethylated CpGs (Fig. \ref{Fig:lessmoms}, purple bars, moments of Eqs. \eqref{eq:proc1}, \eqref{eq:methnumber}) the parameters can still be estimated correctly, however, only with significantly larger sample sizes.
On the other hand, when we only consider the methylation level and the number of methylated Cs per CpG (Fig. \ref{Fig:lessmoms}, brown bars, moments of Eqs. \eqref{eq:methlevel}, \eqref{eq:methnumber}) or the methylation level and the successive (un)methylated CpGs (Fig. \ref{Fig:lessmoms}, gray bars, moments of Eqs. \eqref{eq:methlevel}, \eqref{eq:proc1}, \eqref{eq:proc0}) the GMM can not estimate the real parameters, even for very large sample sizes. 
Fig. \ref{Fig:lessmoms} shows the estimation only for $\mu$, however, the results are very similar for the other parameters and are therefore not shown.
Hence, at least one of the moments derived from the number of successive (un)methylated CpGs (Eq.~\eqref{eq:proc1} or \eqref{eq:proc0}) 
as well as the number of methylated Cs per CpG (Eq.~\eqref{eq:methnumber}) are needed to ensure
identification of the parameters. 

Intuitively, the reason that these moments contain enough information to successfully identify the parameters is that due to the neighborhood dependencies, the average number of consecutive (un)methylated CpGs is a good indicator for the strength of the neighborhood dependence.
Furthermore, since each CpG is influenced by its neighboring CpGs, each CpG in general may have a different average number of methylated Cs. 
The other moments are less informative.  
The average methylation level in Eq. \eqref{eq:methlevel}, for example, gives no hint about the distribution of   methylation, i.e. if it is spread uniformly over all CpGs or only concentrates on certain areas.
On the other hand, once identification is ensured,   additional information from  such moments helps to estimate the parameters more accurately for smaller sample sizes.
For $100-1000$ sample patterns,
which is the order of magnitude for the hairpin bisulfite sequencing data considered later,
  all moments should be considered to achieve an accurate estimation.

We also perform estimations for different parameter sets with  stronger/weaker dependencies, higher/lower methylation efficiencies and combinations thereof.
The results are in agreement with the results   in Fig.~\ref{Fig:compGMMMLE} and \ref{Fig:lessmoms}, i.e., GMM and MLE show a similar accuracy if the sample size is at least of the order of hundreds and also the moment subsets
comparison gives very similar results.
We therefore do not present detailed results for these parameter sets. 

Finally, we apply the GMM to the hairpin bisulfite sequencing  data set 
from mouse ESCs in \cite{arand2012vivo}.
During hairpin bisulfite sequencing, the two DNA strands are linked together covalently such that the methylation status of both strands can be measured simultaneously \cite{laird2004hairpin}.
Our data sets consist of data for single copy genes, which occur only once in the genome, as well as repetitive elements, which occur in multiple copies over the whole genome.
For single copy genes, we have data for Afp (5 CpGs) and Tex13 (10 CpGs).
For the repetetive elements, the data stems from IAP (intracisternal A particle; 6 CpGs), L1 (Long interspersed nuclear elements; 7 CpGs) and mSat (major satellite; 3 CpGs).
We focus on Dnmt1 KO data, i.e. only Dnmt 3a/b is active, since previous findings suggest, that in general only Dnmt 3a/b shows a dependence on the left neighbor, while Dnmt1 acts independent of the neighborhood \cite{luck2019hidden}.

 Since the number of possible states grows exponentially with the number of CpGs, i.e. for $L$ CpGs there are $4^L$ possible states, the numerical solution is no longer feasible, due to large memory requirements for more than 5 CpGs.
We therefore estimate the theoretical moments via MC sampling of the model.
Due to finite size effects and statistical inaccuracies these moments are not exact anymore.
In order to have an estimate for these variations we compute the confidence interval
\begin{equation}
\bar{m}_q\pm 1.96\cdot\sqrt{\frac{S_q^2}{N}},
\end{equation}
where 1.96  is the approximate value of the corresponding percentile point of the normal distribution   for a confidence level of 95\%, $\bar{m}$ and $S^2$ are the sample mean and variance of the quantities in Eqs. \eqref{eq:methlevel}-\eqref{eq:varnumber} for a sample size of $N$.
We find that for $N=1000$ the relative width of the confidence interval is $\leq 0.1$ for all moments and   parameter sets and use this sample size for the approximation of the theoretical moments.

Since we have only one data set for each locus available, we use bootstrapping to generate $25$ samples and again calculate the mean and standard deviations of the estimators.
The results for all available loci are summarized in Tab.~\ref{Tab:GMMMLE}.
Note that the standard deviations are rather large due to multiple reasons.
First of all, the aforementioned variability in the (MC sampled) theoretical moments leads to a variability in the estimates as well.
Furthermore, we use the same parameters for all CpGs. Hence, the results represent the average dependency and methylation efficiency at this position (spanning several CpGs).
Introducing separate parameters for each CpG results in $4L$ parameters and may lead to identifiability problems, due to the in general low coverage.
For the artificial data considered above, we used the same parameters to generate the data, such that the parameters for each CpGs were indeed identical in this case.
 Finally, the number of pattern samples
 that can be considered for the estimation is often very small when considering all CpGs, since often the methylation state for one (or more) of the CpGs is missing, such that we have to omit the whole measurement  (see the second column in Tab. \ref{Tab:GMMMLE} for detailed numbers).
Nevertheless, the results are in good agreement with the previous findings, i.e., for Dnmt 3a/b there is, in general, only a dependence on the left neighbor.

\renewcommand{\arraystretch}{1.25}
\begin{table}[tb]
\begin{center}
\caption{Mean and standard deviations for GMM for BS-seq hairpin data from different loci, obtained from 25 bootstrap samples. \label{Tab:GMMMLE}}
\begin{tabular}{|c|c|c|c|c|c|}
\hline 
~Locus~  & $N$ & $\mu$ & $\psi_L$ & $\psi_R$ & $\tau$ \\ 
\hline 
mSat  & ~$1191$~ & $0.3278 \pm 0.1836$ & $0.2388 \pm 0.1784$ & $0.9624 \pm 0.0743 $ &  $0.0069 \pm 0.0157$ \\
\hline
Afp & $134$ & $0.3700 \pm 0.3254$ & $0.4357 \pm 0.3126 $ &  $0.5254 \pm 0.2833$ & $0.4745 \pm 0.2598$ \\
\hline
IAP & $182$ & $0.5736 \pm 0.1611$ & $0.3868 \pm 0.2738$ &  $0.9388 \pm 0.1044$ & $0.0264 \pm 0.0356$\\
\hline
L1 & $147$ & ~$0.6147 \pm 0.2751$~ & ~$0.9443 \pm 0.1968$~ & ~$0.9596 \pm 0.1959$~ & ~$0.0401 \pm 0.1720$~ \\  
\hline 
Tex13 & $394$ & $0.7039 \pm 0.3474$ & $0.5990 \pm 0.3753$ & $0.9688 \pm 0.0709$ &  $0.9626 \pm 0.0984$\\
\hline
\end{tabular}
\end{center}
\end{table}

\section{Conclusion}
In this paper we   presented a likelihood-free parameter estimation method for DNA methylation models, 
which are based on a mechanistic 
description of methylation pattern formation. 
Our estimation approach is  based on the  generalized method of moments (GMM)
and avoids the expensive (or even infeasible) computation of likelihoods.
We proposed a suitable set of moments and investigated which of these moments are most informative for identification of the parameters.
It turns out that only a minimal set of moments (number of methylated cytosines (C) for each CpG and number of consecutive methylated Cs) is sufficient in order to identify and successfully estimate the parameters.
For a small number of reads, however, information from all defined moments are needed.
The accuracy of our GMM-based approach is comparable to that of maximum likelihood estimation  (MLE) but for longer patterns only the GMM is feasible.
 
 Although the model's moments can be
 estimated by Monte-Carlo sampling,
 a numerical approach to compute the moments without calculating the full underlying distribution is desirable.
As future work, we plan to derive  moment equations that  allow a fast numerical
computation of the statistical moments.
This would allow to   obtain  
 accurate estimates very efficiently
 also   in the case of long methylation patterns and to estimate parameters on a whole-genome scale.
\bibliographystyle{splncs03}

\end{document}